\documentclass[journal=nalefd,manuscript=letter]{achemso}

\usepackage{chemformula} 
\usepackage[T1]{fontenc} % Use modern font encodings
\usepackage[version=4]{mhchem}
\newcommand{\tc}{$T_\mathrm{c}$}

\author{Xiaoyu Wang}
\affiliation{Department of Chemistry, University at Buffalo, Buffalo, NY, 14261, USA}
\author{Warren E. Pickett}
\affiliation{Department of Physics and Astronomy, University of California Davis, Davis, CA, 95616, USA}
\author{Matthew N. Julian}
\affiliation{Intellectual Ventures, Bellevue, WA, 98007, USA}
\author{Rohit P. Prasankumar}
\affiliation{Intellectual Ventures, Bellevue, WA, 98007, USA}
\author{Eva Zurek}
\email{ezurek@buffalo.edu}
\affiliation{Department of Chemistry, University at Buffalo, Buffalo, NY, 14261, USA}
\title{Single layer clathrane: A potential superconducting two-dimensional (2D) hydrogenated metal borocarbide}


\begin{document}

\begin{tocentry}

\includegraphics[width=8.25cm,height=4.45cm]{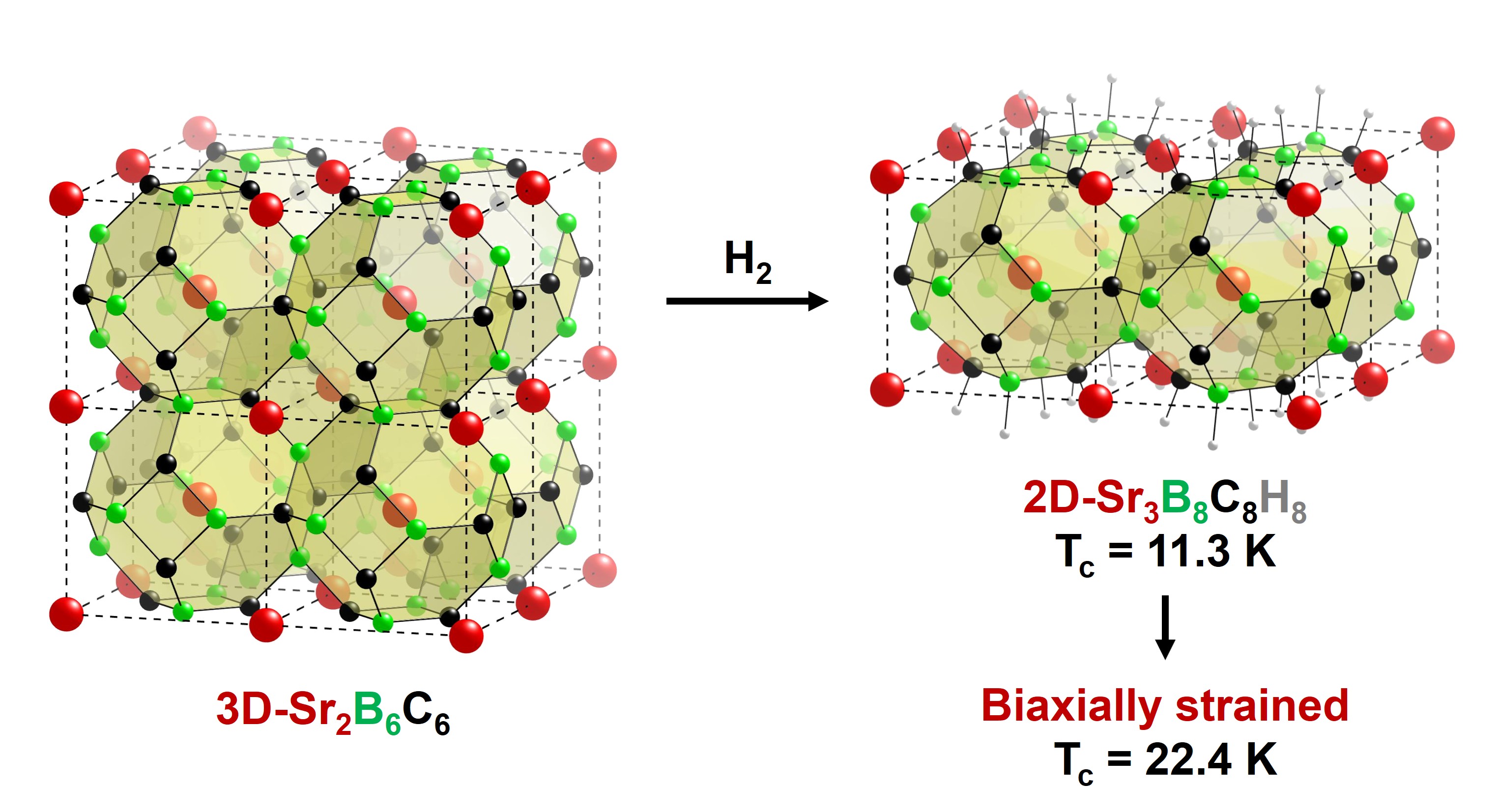}

\end{tocentry}


\newpage
\begin{abstract}
  We propose a new family of two-dimensional (2D) metal-borocarbide clathrane superconductors, derived from three-dimensional (3D) MM$^\prime$B$_6$C$_6$ clathrates. First-principles calculations reveal that hydrogen passivation and surface metal decoration stabilize the M$_2$M$^\prime$B$_8$C$_8$H$_8$ monolayers. These 2D systems exhibit tunable superconductivity governed by hole concentration, structural anisotropy, and electron-phonon coupling. We find that in-plane anisotropy competes with superconductivity, reducing \tc\ despite favorable doping. Biaxial strain mitigates this anisotropy, enhances Fermi surface nesting, and increases \tc\ by an average of 15.5~K. For example, the \tc\ of Sr$_3$B$_8$C$_8$H$_8$ is predicted to increase from 11.3~K to 22.2~K with strain engineering. These findings identify 2D clathranes as promising, strain-tunable superconductors and highlight design principles for optimizing low-dimensional superconducting materials.
\end{abstract}

\newpage

%%%%%%% Introduction
Nanostructuring, for example by decreasing dimensionality, harnessing substrate-proximity effects, or strain engineering can influence a material's superconducting properties~\cite{Lang2024,Krogstrup2015_NatMater,Ruf2021_NatCommun}. Nearly a century ago the first studies on how dimensionality reduction affects superconductivity were performed on Pb and Sn~\cite{Shalnikov1938_Nature}. By now, highly crystalline two-dimensional (2D) systems can be investigated~\cite{Saito2016_NatRevMater}. The first demonstration of superconductivity in the ultimate 2D limit of a single atomic layer yielded a critical temperature (\tc) of 1.8~K for Pb grown epitaxially on Si(111)~\cite{Zhang2010_NatPhys}.   Beyond fundamental interest, 2D superconductors are attractive for device miniaturization, enabling applications in quantum information science, nanoscale circuitry, and gate–tunable superconducting devices~\cite{Liu2019_NatRevMater,Shabani2016_PRB,Gupta2023_NatCommun}.

The superconducting properties of various 2D systems, including alkali-metal-intercalated few-layer graphene~\cite{Xue2012_JACS,Li2013_ApplPhysLett,Tiwari2017_JPCM}, NbSe$_2$~\cite{Frindt1972_PRL,Ugeda2016_NatPhys,Tsen2016_NatPhys,Xi2016_NatPhys}, 2H-NbS$_2$~\cite{wang2021superconducting} and TiSe$_2$~\cite{Li2016_Nature} have been reported. The measured \tc s are sensitive to the nature of the intercalant, competition between superconductivity and charge density waves (CDW), layer thickness, and external electric fields. 
Density functional theory (DFT) studies predicted that the \tc\ of bulk \ce{LiC6} increases from 0.9~K to 8~K in the monolayer~\cite{Profeta2012_NatPhys}, yielding a \tc\ of 18~K for monolayer \ce{Li2C6}~\cite{Yang2024_AdvFuncMater}. The influence of layer thickness, and magnetic and CDW order on the superconducting properties of NbSe$_2$ were investigated~\cite{Zheng2018_PRB,Lian2018_NanoLett}. Strain-tunable 2D superconductors were predicted, including monolayer \ce{W2N3}~\cite{Campi2021_NanoLett} and Janus monolayers of MoSH~\cite{Liu2022_PRB,Ku2023PRB}, WSeH and WSH~\cite{Seeyangnok2024_2DMater,Seeyangnok2024_PRB}.

Boron-based compounds have also attracted considerable interest. While bulk MgB$_2$ exhibits a \tc\ of $\sim$40~K~\cite{Choi2002_Nature}, the \tc\ of a monolayer was predicted to be $\sim$20~K, which could be enhanced to over 50~K under biaxial strain~\cite{Bekaert2017_PRB},  further increased to 67~K via hydrogenation and even 100~K with additional strain engineering~\cite{Bekaert2019_PhysRevLett}. 
Other theoretically studied boron-rich systems include monolayers of B$_2$C~\cite{Dai2012_Nanoscale},  \ce{B2O}\cite{Yan2020_npjComputMater}, \ce{TiB3C}, \ce{Ti2B3C2}~\cite{Li2025_ApplSurfSci}, and LiBC~\cite{Modak2021_PRB}. 
Boron-side hydrogenation of LiBC was predicted to increase \tc\ from 70~K to 80~K~\cite{Modak2021_PRB}, while a \ce{TiB3CH2} monolayer was computed to have a  \tc\ of 18.7~K~\cite{Li2025_ApplSurfSci}.  A high-throughput study reported a \tc\ of 22~K for \ce{Mg2B4N2}~\cite{Wines2023_NanoLett}, isostructural to previously predicted \ce{Mg2B4C2} (\tc\ $\sim$47~K~\cite{Singh2022_npjQuantMater}). Though experiments have yet to verify these computations, the borophene allotrope of boron was synthesized~\cite{Ou2021_AdvSci}, with computed \tc s reaching as high as 10–20~K~\cite{Penev2016_NanoLett}. The reactivity of borophene with oxygen hinders such applications. However, recent experiments demonstrated that hydrogen passivation can significantly enhance borophene’s stability~\cite{Li2021_Science}, and subsequent theoretical studies predicted an increase in \tc\ up to 29~K under uniaxial strain~\cite{Soskic2024_NanoLett}. 

Herein, superconductivity in the 2D analogues of a new class of borocarbides is studied via DFT calculations. The compounds \ce{SrB3C3}\cite{Zhu2023_PRRes,Zhu2020_SciAdv} and \ce{LaB3C3}\cite{Strobel2021_AngewChem} have been synthesized in bulk with \ce{SrB3C3} exhibiting a \tc\ of 22~K at 23~GPa~\cite{Zhu2023_PRRes}. 
Subsequent theoretical studies predicted additional superconducting phases including \ce{BaB3C3}~\cite{Wang2021_PRB}, \ce{RbSrB6C6}~\cite{Zhang2022_PRB}, Rb$_{0.8}$Sr$_{1.2}$B$_6$C$_6$~\cite{Gai2022_PRB}, \ce{KPbB6C6}~\cite{Geng2023_JACS}, \ce{CsBaB6C6}~\cite{DiCataldo2022_PRB}, \ce{RbYbB6C6}~\cite{Duan2024_PRB},  \ce{SrNH4B6C6} and \ce{PbNH4B6C6}~\cite{Sun2024_CommunPhys} with \tc s as high as 115~K~\cite{Sun2024_CommunPhys}. Anharmonicity was shown to effect the \tc\ of \ce{KPbB6C6}~\cite{Zhao2025_arxiv}. Structurally, MM$^\prime$\ce{B$_6$C$_6$} clathrates resemble diamond -- both are covalent solids built on $sp^3$ frameworks~\cite{Geng2023_JACS}. The synthesis conditions for compounds such as \ce{SrB6C6} (50–60 GPa and 2500 K)\cite{Zhu2020_SciAdv,Zhu2023_PRRes} closely resemble those required for diamond formation (typically $>$~15~GPa and 1200~K)\cite{Zhu2020_Matter,Xie2017_JACS,Khaliullin2011_NatMater}. 
However, while diamond has a 2D layered counterpart known as diamane, which can be experimentally realized from bilayer graphene under pressure and/or surface passivation with hydrogen or fluorine\cite{Sorokin2021_NanoLett}, the 2D analogue of \ce{M$_2$B$_6$C$_6$}—a metal-borocarbide “clathrane”—has not yet been explored.
Below we investigate whether such a clathrane monolayer is theoretically viable, and examine its superconducting properties.

\begin{figure}[!htbp]
    \centering
    \includegraphics[width=0.8\linewidth]{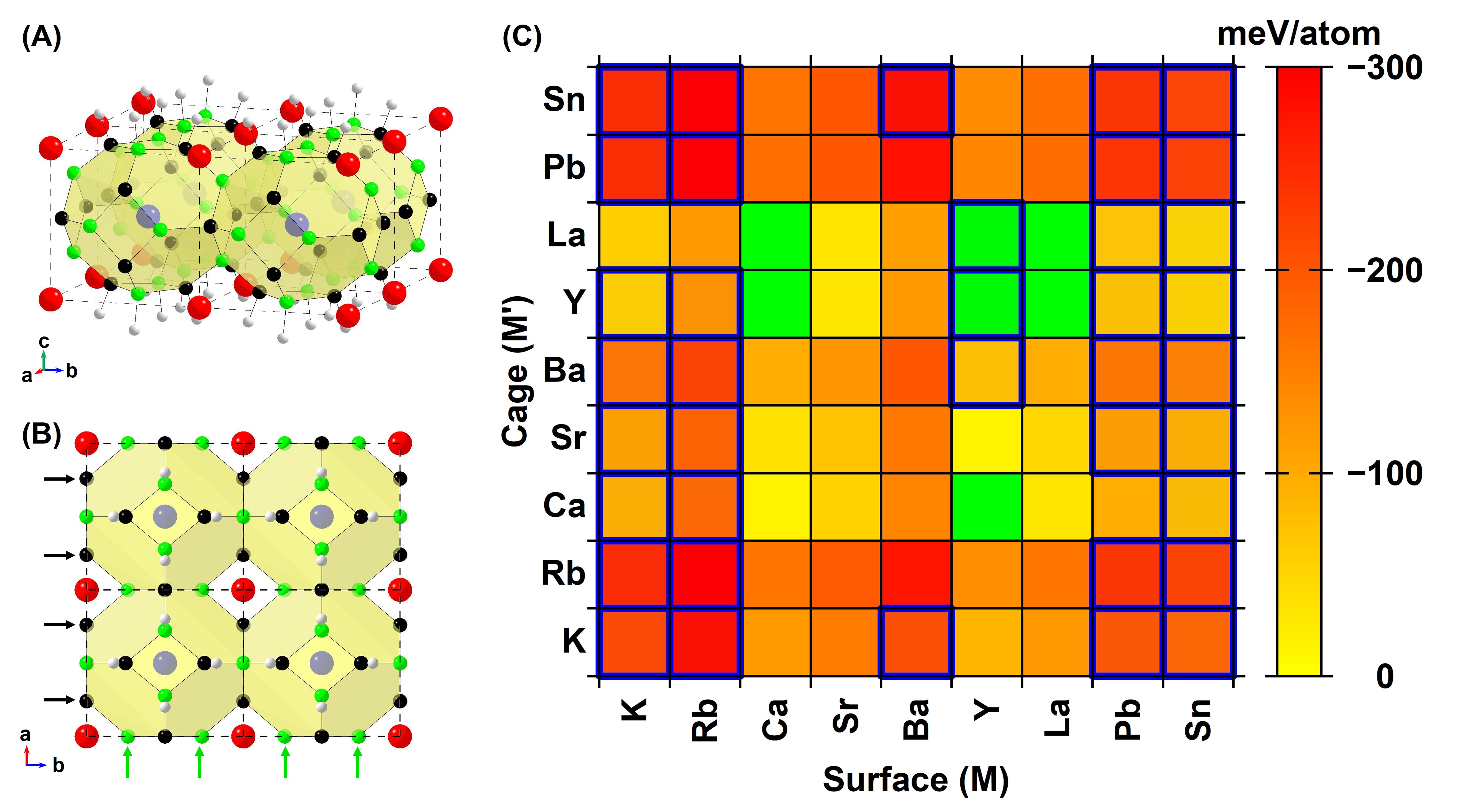}
    \caption{(A) Supercell of the 2D hydrogenated metal borocarbides (M$_2$M$^\prime$B$_8$C$_8$H$_8$). The color of the surface metals is red, cage metals are gray, boron is green, carbon is black, and hydrogen is white. (B) Top view of the structure; the equatorial boron/carbon atoms are marked with green/black arrows. (C) The energy ($\Delta E_\textsc{EXF}$) associated with the formation of the 2D structures along with solid M$^\prime$, $\alpha$-boron, and graphite via hydrogenation of the 3D structure. The green color shows regions where 3D structures are thermodynamically preferred (Ca$_2$Y, Ca$_2$La, Y$_2$Ca, Y$_3$, Y$_2$La, La$_2$Y, and La$_3$), otherwise the color is scaled to the energy difference. Dynamically unstable structures are enclosed with blue squares.}
    \label{fig:1}
\end{figure}

%%%%%%% Structure and stability
Our model, illustrated in Figure~\ref{fig:1}(A,B), features a single-layer of the 3D clathrate with metal atoms positioned both at the center of the cage and on the surface within a half-cage configuration. 
Similar to diamane~\cite{Piazza2020_Carbon} and borophene~\cite{Li2021_Science}, surface passivation is essential for dynamic stability. Various termination schemes were evaluated—no passivation, bridge-site passivation with oxygen or sulfur, halogen termination, and hydrogenation—and only hydrogenation yielded local minima. 
The presence of surface metal atoms was also found to be critical for dynamic stability. 
By combining appropriate surface passivation and metal incorporation we arrived at a stable composition with M$_2$M$^\prime$B$_8$C$_8$H$_8$ stoichiometry, where M represents the surface metal and M$^\prime$ denotes the encapsulated metal cation (Figure~\ref{fig:1}A). 
The choice of metals was guided by size compatibility with the B$_6$C$_6$ cage,  including K, Rb, Ca, Sr, Ba, Y, La, Sn, and Pb, following previous theoretical results~\cite{Geng2023_JACS}.

Computations investigated the thermodynamic conditions required to stabilize the 3D clathrates~\cite{Zhu2020_SciAdv,Geng2023_JACS, DiCataldo2022_PRB}, and synthesized \ce{SrB3C3} was quenched to 1~atm, persisting in an inert atmosphere but degrading after exposure to moisture~\cite{Zhu2020_SciAdv}. For the clathranes, the thermodynamic stability (at zero temperature and for static nuclei) was assessed by calculating the exfoliation energy ($\Delta E_\textsc{EXF}$) defined as the energy of the products minus the reactants for the reaction:
\begin{equation*}
\text{MM}^\prime \text{B}_6 \text{C}_{6\text{(s,3D)}} + 4 \text{H}_{2\text{(g)}} \rightarrow  \text{M}_2\text{M}^\prime \text{B}_8\text{C}_8 \text{H}_{8\text{(s,2D)}} + \text{M}^\prime_\text{(s)} + 4 \text{B}_\text{(s)} + 4 \text{C}_\text{(s)}, \label{eq:rxn}
\end{equation*}
where B, C, and M$^\prime$ correspond to the 1~atm stable phases of $\alpha$-boron, graphite, and solid elemental metals in the $fcc$ (K, Ca, Pb), $bcc$ (Rb, Ba), $hcp$ (Sr, Y, La), or $\alpha$-Sn structure, and H$_2$ is modelled by the $P6_3/m$ phase of solid molecular hydrogen~\cite{Gregoryanz2020_MatRadExtreme}. The functional used underestimates the energy of graphite, and inclusion of dispersion is likely to stabilize the products. In many cases, with exceptions such as Sn and Pb, the formation of a binary metal boride or metal carbide would strongly favor the forward reaction (Table S2).
Therefore, the energy estimates presented here represent a lower bound for the thermodynamic driving force favoring hydrogenated 2D slab formation following the exfoliation of the 3D crystal in a hydrogen atmosphere.

In most cases the aforementioned process is exothermic (Figure~\ref{fig:1}C). Exceptions, highlighted in green, involve lanthanum, yttrium and calcium atoms where $\Delta E_\textsc{EXF}$ was predicted to range from 5.2 to 48.0~meV/atom.  The reason for this is likely the enhanced stability of MM$^\prime$\ce{B$_6$C$_6$} compounds that contain trivalent metal atoms, because they are insulators. Notably, $\Delta E_\textsc{EXF}$ for the formation of Sr$_3$B$_8$C$_8$H$_8$ from Sr$_2$B$_6$C$_6$ was —72.3~meV/atom. The tendency for the formation of  \ce{SrB6} and \ce{SrC2} on the products side further suggests that  the forward reaction would be favored. A similar stabilizing effect has been established in diamane~\cite{Sorokin2021_NanoLett}.  Without surface passivation, diamond nanosheets are unstable and reconstruct into multilayer graphene~\cite{Kvashnin2014_NanoLett}. The strong C-H bonds ensure that diamane is resistant to dissociation with activation barriers reaching up to 6~eV\cite{Mortazavi2020_ApplSurfSci}. Our molecular dynamics simulations showed that Sr$_3$B$_8$C$_8$H$_8$ remains intact up to at least 500~K with a simulation length of 10~ps~(Figure~S2 and S3). At above 600~K, we observed the dissociation of the B$_2$C$_2$ squares and reorganize into 5- or 6-member rings.  
These results suggest that SrB$_3$C$_3$, which is thermodynamically unstable at 1~atm and reactive with air, may, upon hydrogenation, resist decomposition due to the strong B-H and C-H bonds formed within Sr$_3$B$_8$C$_8$H$_8$.

Bulk MM$^\prime$B$_6$C$_6$ can be interpreted as a hole-doped compound relative to the ideal sodalite-type \ce{C_6} cage; the number of unoccupied C/B 2$p$ states (i.e., holes) was found to correlate with dynamic stability~\cite{Zhang2022_PRB}. 
For example, in  (Rb,Sr)$_2$B$_6$C$_6$ imaginary phonons emerge when the rubidium concentration exceeds 50\%\cite{Zhang2022_PRB}, and no dynamically stable clathrates were found when both M and M$^\prime$ were alkali metals~\cite{DiCataldo2022_PRB,Geng2023_JACS,Duan2024_PRB}.
For the 2D clathranes no dynamically stable structure was identified when the number of holes exceeded four per formula unit (Figure~\ref{fig:1}C).  Previous studies reported a correlation between dynamic stability and the ionic radius mismatch between the two guest metal species in the 3D clathrates~\cite{Geng2023_JACS}—an effect that also manifests in the 2D analogues--further underscoring the necessity of both surface passivation and the inclusion of surface metal atoms. %Without these the unsaturated $sp^3$ bonds experience an excessive hole concentration destabilizing the lattice.

\begin{figure}[!ht]
    \centering
    \includegraphics[width=1.0\linewidth]{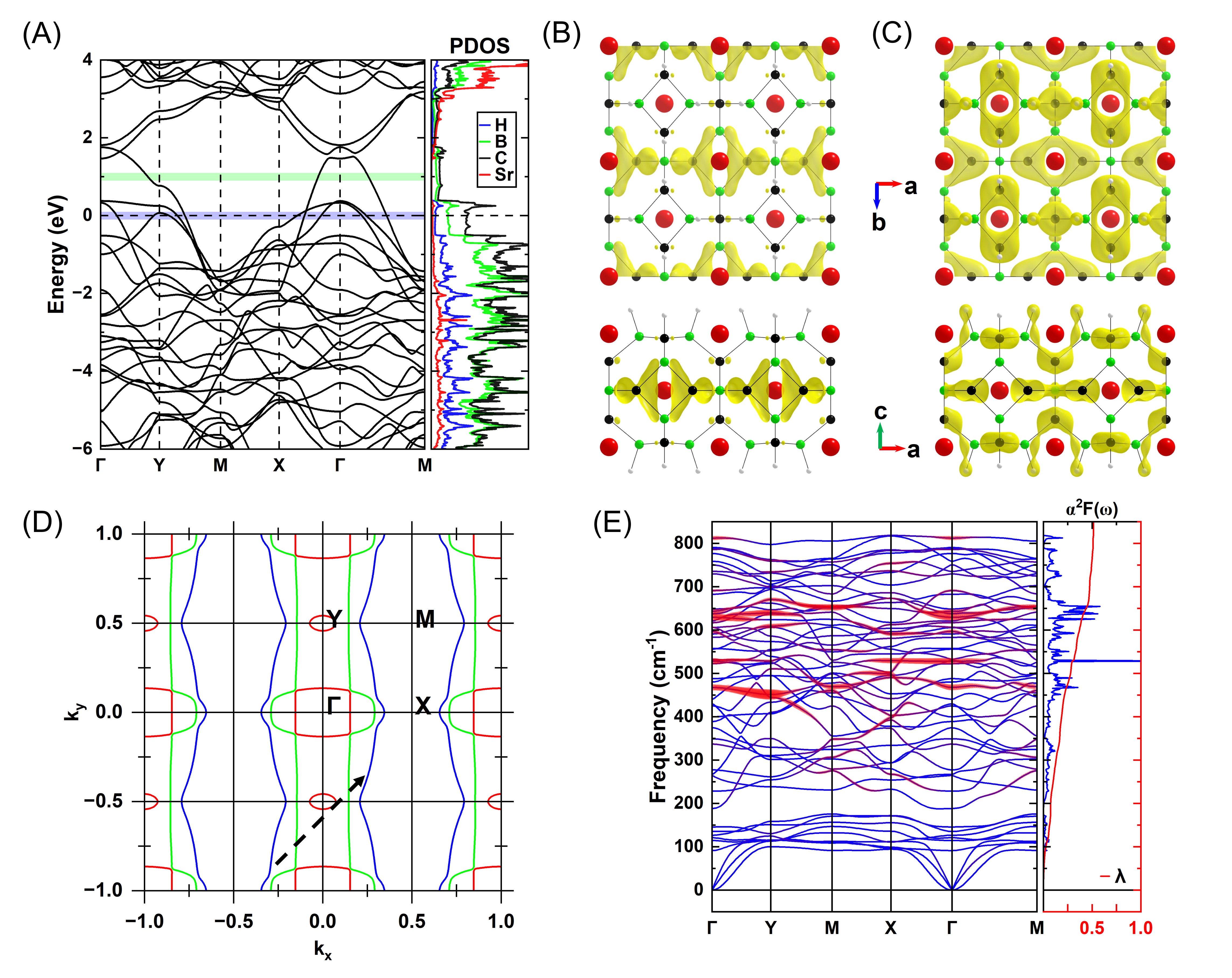}
    \caption{Electronic structure of Sr$_3$B$_8$C$_8$H$_8$. (a) Electronic band structure and atom-projected density of states (PDOS) of Sr$_3$B$_8$C$_8$H$_8$. The Fermi level ($E_\mathrm{F}$) is set to 0~eV. (b) Partial charge density integrated around $E_\mathrm{F}$+1~eV within an energy window of $\pm$0.1~eV, corresponding to the green-shaded region in (a). (c) Partial charge density integrated around $E_\mathrm{F}$ within an energy window of $\pm$0.1~eV, corresponding to the blue-shaded region in (a). The isosurfaces in (b) and (c) enclose 50\% of the total charge within the selected energy windows. Atom color coding is consistent with Figure \ref{fig:1}. (d) Fermi surface of the 2D \ce{Sr3B8C8H8} clathrane. The black dashed arrow illustrates an example of a $\mathbf{q}$-vector at the M point ($q_x = 0.5$, $q_y = 0.5$) that connects nested electronic states on the steepest band. (e) Phonon dispersion, Eliashberg spectral function, $\alpha^2\mathrm{F}(\omega)$, and the integrated electron-phonon coupling constant, $\lambda(\omega)$, within the frequency range of 0-800 cm$^{-1}$. The width of the red lines in the phonon dispersion corresponds to the mode-resolved electron-phonon coupling strength, proportional to $\lambda_{\mathbf{q}\nu} \omega_{\mathbf{q}\nu}$ for each phonon mode $\nu$ at wavevector $\mathbf{q}$. The plot in the full frequency range is available in the SI. }
    \label{fig:3}
\end{figure}

% electronic
Our analysis focuses on \ce{Sr3B8C8H8}, as its parent compound \ce{Sr2B6C6} has been synthesized~\cite{Zhu2020_SciAdv}.
A key distinction between the 2D and 3D structures is the loss of chemical equivalence among the carbon and boron atoms in the clathrane. 
As a result, the conduction band states that are triply-degenerate at the $\Gamma$ point, located $\sim$1.5~eV above the Fermi level ($E_\mathrm{F}$) in \ce{Sr2B6C6}\cite{Zhu2020_SciAdv}, split in the 2D structure, with a total energy separation of 1.5~eV (Figure~\ref{fig:3}A). Compared to the 3D system, the uppermost band shifts from $E_\mathrm{F}$ + 1.5~eV to $E_\mathrm{F}$ + 1.8~eV, while the remaining two (non-degenerate) states are stabilized and become nearly flat, lying $\sim$0.3~eV above $E_\mathrm{F}$.
To better understand the electronic structure of this clathrane, we calculated its partial charge densities within chosen energy windows. The green region in Figure~\ref{fig:3}A, spanning an energy range of $\pm$0.1~eV around $E_\mathrm{F}$ + 1~eV, cuts through the aforementioned dispersive band. Its computed charge density resembles $\pi$-type orbitals centered on unpassivated B/C atoms that point along the  $a$ lattice vector, encompassing the \ce{B2C2} square in the central region of the clathrane layer (Figure~\ref{fig:3}B) with minimal contributions from the surface atoms. 
The projected density of states (PDOS) in this energy range is similar to that of the 3D clathrates, where boron and carbon contribute comparably, and no contribution from hydrogen is evident~\cite{Geng2023_JACS,DiCataldo2022_PRB,Wang2021_PRB}.  Notably, the boron atoms that contribute significantly to the charge density in the green-shaded-region do not at all contribute to the region shaded in blue. As a result, the PDOS near $E_\mathrm{F}$ contains significantly larger carbon character  -- a feature rarely observed in the 3D counterparts. 
Additionally, the hydrogen atoms, which passivate the structure, contribute modestly to the PDOS around $E_\mathrm{F}$. This hydrogen character is evident in the charge density plot, especially the lower panel in Figure~\ref{fig:3}C. Another notable feature is the nearly flat electronic DOS  -- a hallmark of a 2D system -- around $E_\mathrm{F}$, contrasting the nearly parabolic DOS in \ce{Sr2B6C6}.

The three bands that cross $E_\mathrm{F}$ form the 2D Fermi surface of Sr$_3$B$_8$C$_8$H$_8$ (Figure~\ref{fig:3}D), and they can participate in the superconducting mechanism. The Fermi surface is characterized by nearly vertical features located at $k_x \approx 0.1$, which enable favorable Fermi surface nesting across a wide range of $\mathbf{q}$-vectors along the $\Gamma$–$Y$ path. This nesting condition is clearly manifested in the phonon spectrum (Figure~\ref{fig:3}E), where three relatively flat phonon branches near 470, 530,  and 630 cm$^{-1}$ display strong EPC along the $\Gamma$–$Y$ direction, consistent with the three dominant peaks in the Eliashberg spectral function.  Significant EPC-active modes are also found near the $M$ point; the nesting vector $\mathbf{q} = (0.5, 0.5)$ connects sections of the Fermi surface (black arrow in Figure~\ref{fig:3}E), emphasizing the anisotropic nature of the EPC. Significant nesting is also found along the other high-symmetry lines in the Brillouin Zone, though in a very narrow frequency range.

The dominant EPC mechanism in bulk SrB$_3$C$_3$ arises from the interaction between C-2$p$ states and an $E_g$ phonon mode near $\Gamma$, involving out-of-plane displacements of boron atoms leading to distortions of the B$_2$C$_2$ squares~\cite{Wang2021_PRB}. 
We identified a similar vibrational mode in Sr$_3$B$_8$C$_8$H$_8$ at $\sim$640~cm$^{-1}$, however it contributed significantly to EPC only near the zone center, and its contribution rapidly decayed moving away from $\Gamma$. Visualization of the three aforementioned phonon branches with large EPC in the clathrane reveals a Jahn–Teller-like distortion of the B$_2$C$_2$ units, transforming square motifs into rectangles.
The 470~cm$^{-1}$ mode corresponds to distortions of the central $ab$-plane square within the cage; the 630~cm$^{-1}$ mode involves similar distortions at the surface $ab$-plane square, but with additional motion from the hydrogen atoms increasing its frequency. 
The 530~cm$^{-1}$ mode primarily involves distortions of the $ac$-plane square.
Previously, we identified a similar Jahn-Teller mechanism as a dominant contributor to the EPC in many MM$^\prime$B$_6$C$_6$ systems with 2 or 3 holes per formula unit~\cite{Geng2023_JACS}. 
These compounds, such as \ce{KPbB6C6}, typically exhibited phonon softening, however, such softening is absent in Sr$_3$B$_8$C$_8$H$_8$.
This combination of broken degeneracy in the electronic structure and the relatively ``hard'' phonon modes weakens the EPC in Sr$_3$B$_8$C$_8$H$_8$, resulting in a moderate coupling constant of $\lambda = 0.55$ (compared to $\lambda = 0.92$ in \ce{SrB6C6} \cite{Wang2021_PRB}). 
The high-frequency B-H and C-H stretches do not contribute significantly to EPC in Sr$_3$B$_8$C$_8$H$_8$. 
Nonetheless, a logarithmic average phonon frequency, $\omega_{\log}$, of 600~K is obtained, slightly higher than that of the bulk (544~K)\cite{Wang2021_PRB}, reflecting the lack of phonon softening. 
Solving the isotropic Eliashberg equations with $\mu^\star = 0.1$ yields a $T_c$ of 11.3~K, significantly lower than the theoretically predicted $T_c$ of 40~K for bulk Sr$_2$B$_6$C$_6$~\cite{Wang2021_PRB}.

%We now examine how variations in the identity of the metal atoms influence the structural, electronic, and superconducting properties of the 2D clathranes. 
The $T_c$s of binary MM$^\prime$B$_6$C$_6$ compounds correlate with the level of hole doping, with the highest values obtained for metals with an average charge of +1.5~\cite{Wang2021_PRB,Geng2023_JACS,DiCataldo2022_PRB,Duan2024_PRB}.  While the valence and radius of the metal atoms are important for the 2D clathranes, an additional distinguishing feature—also intimately linked to hole doping and ionic size—is the in-plane structural anisotropy. 
This anisotropy arises from the inequivalent in-plane atomic configurations: carbon atoms align along the equatorial $a$-axis, while boron atoms align along the $b$-axis (Figure~\ref{fig:1}B). 
Similar anisotropy is present in 2D \ce{B2O} and hydrogenated $\beta_{12}$-borophene\cite{Yan2020_npjComputMater,Soskic2024_NanoLett}.
To capture this effect, we fully optimized the in-plane lattice parameters $a$ and $b$, imposing zero stress along both directions. 
The resulting difference in lattice constants ($b - a$) serves as a metric for quantifying the degree of in-plane anisotropy.
We find that structures with higher hole concentration generally exhibit larger anisotropy (Figure S1). 
In compositions such as La$_3$, La$_2$Y, Y$_2$La, and Y$_3$, where all B/C 2$p$ orbitals are fully occupied and any excess electrons reside in the metal $d$ states, the lattice distortion is minimal, with anisotropy values typically below 0.1~\r{A}. 
In contrast, 1~$h^+$ structures exhibit anisotropies ranging from 0.05–0.25~\r{A}, 2$h^+$ from 0.12–0.39~\r{A}, and 3$h^+$ from 0.21–0.52~\r{A}. 
Among clathranes with the same hole concentration, those incorporating metals with smaller ionic radii exhibit larger anisotropy (Figure~S1), indicating a synergistic effect between doping level and cation size. 
Importantly, this anisotropy competes with superconductivity, suggesting a trade-off between lattice distortion and optimal EPC.

\begin{figure}[!ht]
    \centering
    \includegraphics[width=1.0\linewidth]{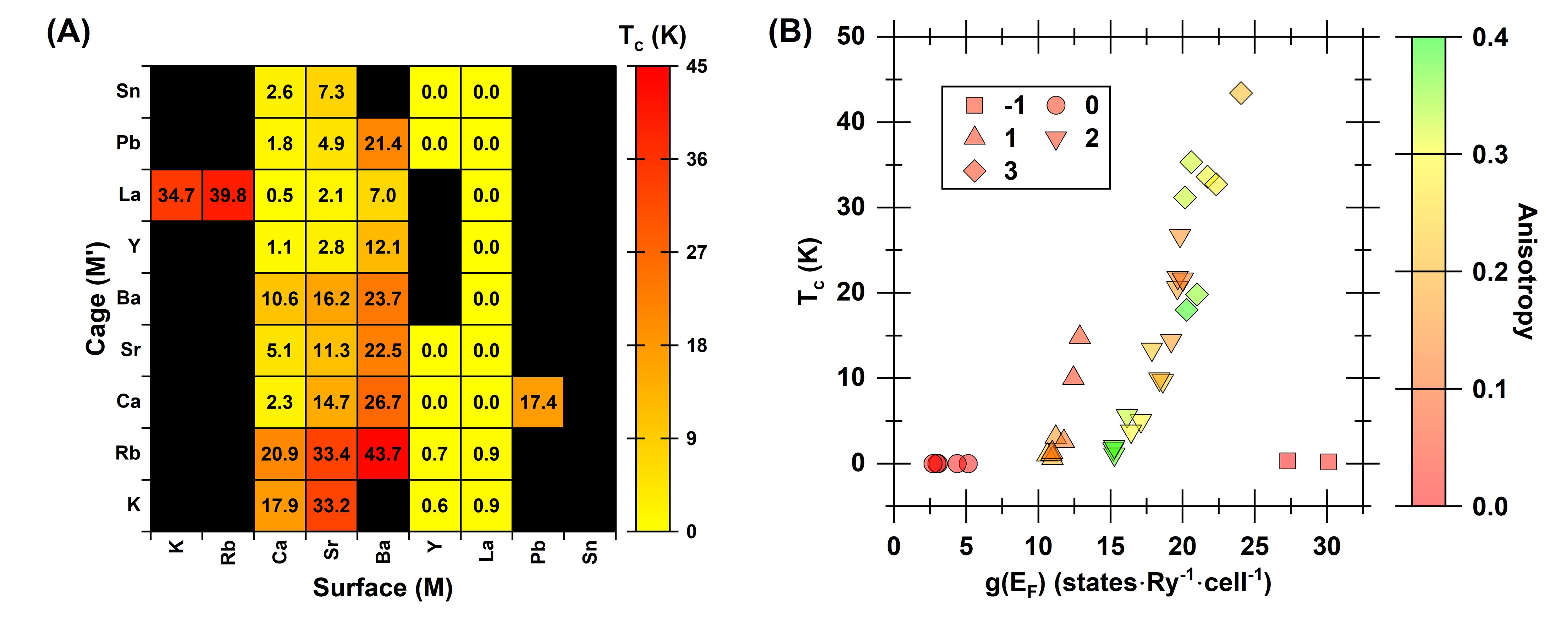}
    \caption{(a) Summary of the superconducting transition temperatures ($T_c$s) for various \ce{M$_2$M$^\prime$B$_8$C$_8$H$_8$} compositions under zero external stress, i.e., fully relaxed in-plane lattice constants. The $T_c$s are calculated using Eliashberg theory with a $\mu^\star$ of 0.1; for systems with $T_c < 10$ K, values are estimated using the Allen-Dynes modified McMillan formula. (b) Correlation between $T_c$, the density of states at the Fermi level, $g(E_\textsc{F})$, and the in-plane structural anisotropy in \AA{} (quantified by the lattice constant difference $b - a$). The inset shows the symbols employed for systems with the listed number of holes, or with an excess electron (denoted by -1).}
    \label{fig:4}
\end{figure}

For 3D clathrates the $T_c$s within a given hole-doping group are relatively consistent, ranging from 30–44 K for 1~$h^+$ systems, 40-54~K for 2$h^+$, and 72-88~K for 3$h^+$~\cite{Geng2023_JACS}. 
In contrast, 2D clathranes show a broader and more irregular $T_c$ spread (Figure \ref{fig:4}A), suggesting additional factors are at play. 
For example, within the 1~$h^+$ group, $T_c$ ranges from 0-15~K; for 2~$h^+$, from 1-27~K; and for 3~$h^+$, from 18-43~K. 
Figure~\ref{fig:4}B illustrates the $T_c$, in-plane anisotropy, and the electronic density of states at the Fermi level ($g(E_\textsc{F}$)).
In general, an increased number of holes leads to a higher $g(E_\textsc{F}$), consistent with enhanced metallicity. 
An exception is seen in electron-doped species (–1 hole, e.g., La$_3$, La$_2$Y, Y$_2$La, Y$_3$), which exhibit the highest $g(E_\textsc{F}$) from the metal $d$-states. Structures with zero holes are nominally semiconducting, although in some cases band overlap results in a small but nonzero $g(E_\textsc{F})$.
Notably, structures with anomalously low \tc—despite having favorable doping levels—exhibit larger in-plane anisotropy compared to other members of the same group, reinforcing the conclusion that in-plane anisotropy suppresses superconductivity in 2D clathranes, disrupting the otherwise robust correlation between hole doping and \tc\ observed in their 3D counterparts.

\begin{figure}
    \centering
    \includegraphics[width=0.75\linewidth]{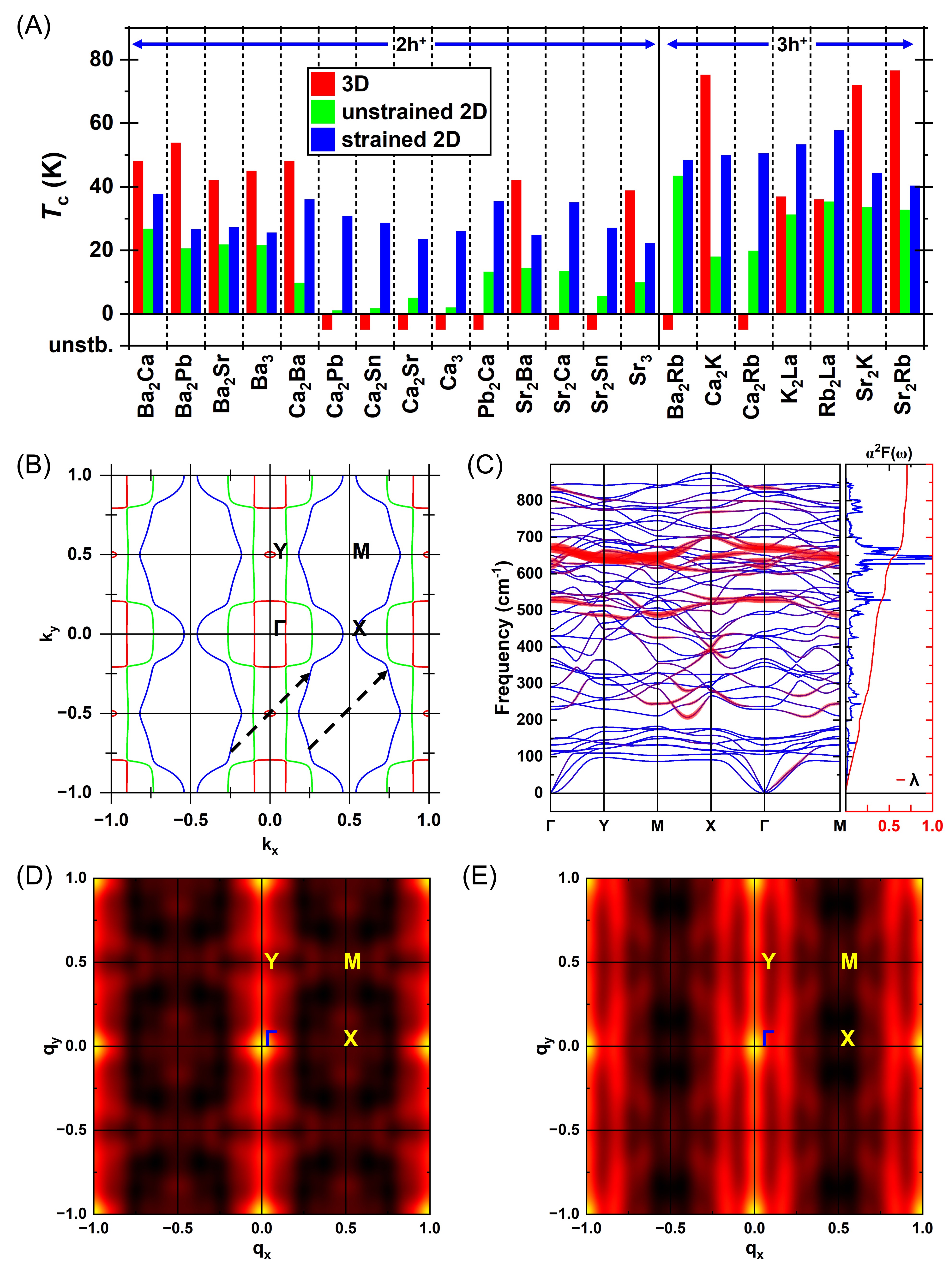}
    \caption{(a) Comparative summary of $T_c$ in 3D bulk  \ce{MM$^\prime$B$_6$C$_6$}, unstrained 2D  \ce{M$_2$M$^\prime$B$_8$C$_8$H$_8$}, and biaxial strained 2D  \ce{M$_2$M$^\prime$B$_8$C$_8$H$_8$}. The $T_c$s are calculated using Eliashberg theory with a $\mu^\star$ of 0.1; for systems with $T_c < 10$ K, values are estimated using the Allen-Dynes modified McMillan formula. \ce{Ca$_2$Pb}, \ce{Ca$_2$Sn}, \ce{Ca$_2$Sr}, \ce{Ca$_3$}, \ce{Pb$_2$Ca}, \ce{Sr$_2$Ca}, \ce{Sr$_2$Sn}, \ce{Ba$_2$Rb}, and \ce{Ca$_2$Rb} are dynamically unstable in 3D but stable in the 2D analogues. (b) Fermi surface of biaxially strained 2D \ce{Sr3B8C8H8}. (c) Phonon dispersion, Eliashberg spectral function, $\alpha^2\mathrm{F}(\omega)$, and the integrated electron-phonon coupling constant, $\lambda(\omega)$, within the frequency range of 0-900 cm$^{-1}$. The width of the red lines in the phonon dispersion corresponds to the mode-resolved electron-phonon coupling strength, proportional to $\lambda_{\mathbf{q}\nu} \omega_{\mathbf{q}\nu}$ for each phonon mode $\nu$ at wavevector $\mathbf{q}$. The plot in the full frequency range is available in the SI. (d) Two-dimensional nesting function, $\chi(\mathbf{q})$, of unstrained \ce{Sr3B8C8H8}. (e) Nesting function of biaxially strained \ce{Sr3B8C8H8}.}
    \label{fig:strained}
\end{figure}

While the relationship between anisotropy and superconductivity likely involves multiple intertwined effects, it is clear that larger in-plane anisotropy correlates with suppressed $T_c$.  To determine if a physical approach mitigating the anisotropy by biaxial strain could enhance \tc , computations were performed that restored the in-plane lattice constants to those of the corresponding 3D clathrates by simultaneously elongating the $a$ and contracting the $b$ lattice vector.
We focused on compositions with 2$h^+$ and 3$h^+$, since their $T_c$s were typically higher (Figure~\ref{fig:4}B).
Comparing $T_c$ across the 3D clathrates, unstrained 2D clathranes, and biaxially strained clathranes revealed that the 3D phases generally exhibit the highest $T_c$, with exceptions such as \ce{K$_2$La} and \ce{Rb$_2$La} (Figure~\ref{fig:strained}A). 
The degeneracy breaking in the 2D structures weakens the EPC, accounting for the lower $T_c$ in most cases. 
Conversely, some dynamically unstable 3D frameworks possess stable clathrane analogues—particularly those containing Ca and Sn.
Biaxial strain enhances $T_c$ relative to the fully optimized systems, with an average increase of 15.5~K (Figure \ref{fig:strained}A), e.g., the $T_c$ of \ce{Sr$_3$B$_8$C$_8$H$_8$} increases from 11.3~K to 22.2~K upon straining. 
This enhancement can be traced to subtle modifications in the electronic band structure: elongation of the $a$ axis results in flatter bands along $\Gamma-X$ and steeper bands along $Y-M$, shifting the Fermi crossing points closer to $X$ and $Y$, respectively (Figure~\ref{fig:strained}B), with concomitant changes in the EPC-active modes (Figure~\ref{fig:strained}C). 
Strain enhances the nesting along the $\Gamma$–M direction, as evidenced by the 2D nesting functions (c.f.\ Figure~\ref{fig:strained}C and Figure~\ref{fig:strained}D).
These improved nesting features activate the same EPC modes as in the unstrained case, but with stronger contributions extending from the $\Gamma$–Y direction to a broader range of $\mathbf{q}$-vectors along $\Gamma$–Y–M and $\Gamma$–M (Figure~\ref{fig:strained}B), increasing $\lambda$ from 0.55 to 0.71. 
However, this enhancement comes with a trade-off: low-frequency phonons around 200~cm$^{-1}$ become EPC-active due to mode softening along $\Gamma$-M and $M$-$X$. These phonon modes involve out-of-plane displacements of boron atoms, following the $E_g$ character described for 3D \ce{Sr2B6C6}\cite{Wang2021_PRB} and decreasing $\omega_{\log}$ from 600.3~K to 531.4~K.

%%CONCLUSION
In conclusion, a family of 2D metal-borocarbide clathranes, derived from 3D MM$^\prime$B$_6$C$_6$ clathrate counterparts, some that have recently been synthesized~\cite{Zhu2020_SciAdv,Strobel2021_AngewChem,Zhu2023_PRRes} was proposed.  DFT calculations demonstrated that dynamic and thermodynamic stability of the 2D analogues can be achieved via appropriate surface metal decoration and hydrogen passivation, yielding stable M$_2$M$^\prime$B$_8$C$_8$H$_8$ compositions. 
The Fermi level and hole concentration can be tuned by varying the metal atoms, enabling metallicity and superconductivity. %We identified a correlation between hole concentration, in-plane structural anisotropy and $T_c$. 
Unlike their 3D analogues, the 2D clathranes exhibit in-plane anisotropy, which competes with, and in some cases suppresses superconductivity. 
Clathranes with high anisotropy typically show reduced EPC and lower $T_c$s, despite favorable doping levels, whereas biaxial strain reduces anisotropy, and enhances superconducting properties. 
Strain engineering increased $T_c$ by an average of 15.5~K, with Sr$_3$B$_8$C$_8$H$_8$ exhibiting a $T_c$ enhancement from 11.3~K to 22.2~K. 
This study establishes 2D clathranes as a promising platform for tunable superconductivity, and highlights the interplay between hole doping, lattice anisotropy, and electron-phonon interactions. 
%Our findings present design principles guiding future synthesis and optimization of low-dimensional superconductors.

\begin{acknowledgement}
This work was supported by the Deep Science Fund of Intellectual Ventures and NSF award DMR-2119065. Calculations were performed at the Center for Computational Research  (http://hdl.handle.net/10477/79221) at SUNY Buffalo. We thank Timothy Strobel and Stu Wolf for fruitful discussions.
\end{acknowledgement}

\begin{suppinfo}
The Supporting Information is available free of charge on the ACS Publication website. It includes full computational details, extra analysis on thermodynamics, projected phonon linewidths and Eliashberg spectral functions, phonon band structures and density of states, electronic structure analysis, trajectories of molecular dynamics runs, and structural parameters. 
\end{suppinfo}

\bibliography{reference}

\end{document}